\documentclass[aps,prl,reprint,superscriptaddress]{revtex4-1} 

\usepackage{amssymb}
\usepackage{amsmath}
\usepackage{graphicx}
\usepackage{dcolumn}
\usepackage{epstopdf}

\begin{document}

\title{Imaging the Anisotropic Nonlinear Meissner Effect in Nodal YBa$_{2}$Cu$_{3}$O$_{7-\delta}$ Thin-Film Superconductors}

\author{Alexander P. Zhuravel}
\affiliation{B. Verkin Institute for Low Temperature Physics and
Engineering, National Academy of Sciences of Ukraine, UA-61103 Kharkov, Ukraine}

\author{B. G. Ghamsari}
\affiliation{CNAM, Physics Department, University of Maryland,
College Park, Maryland USA 20742-4111}

\author{C. Kurter}
\affiliation{CNAM, Physics Department, University of Maryland,
College Park, Maryland USA 20742-4111}

\author{P. Jung}
\affiliation{Physikalisches Institut and DFG-Center for Functional
Nanostructures (CFN), Karlsruhe Institute of Technology, DE-76128
Karlsruhe, Germany}

\author{S. Remillard}
\affiliation{Physics Department, Hope College, 27 Graves Place, Holland, MI USA
49422}

\author{J. Abrahams}
\affiliation{CNAM, Physics Department, University of Maryland,
College Park, Maryland USA 20742-4111}

\author{A. V. Lukashenko}
\affiliation{Physikalisches Institut and DFG-Center for Functional
Nanostructures (CFN), Karlsruhe Institute of Technology, DE-76128
Karlsruhe, Germany}

\author{Alexey V. Ustinov}
\affiliation{Physikalisches Institut and DFG-Center for Functional
Nanostructures (CFN), Karlsruhe Institute of Technology, DE-76128
Karlsruhe, Germany}

\author{Steven M. Anlage}
\affiliation{CNAM, Physics Department, University of Maryland,
College Park, Maryland USA 20742-4111}
\affiliation{Physikalisches
Institut and DFG-Center for Functional Nanostructures (CFN),
Karlsruhe Institute of Technology, DE-76128 Karlsruhe, Germany}

\begin{abstract}
 We have directly imaged the anisotropic nonlinear Meissner effect in an unconventional superconductor through the nonlinear electrodynamic response of both (bulk) gap nodes and (surface) Andreev bound states. A superconducting thin film is patterned into
a compact self-resonant spiral structure, excited near resonance in
the radio-frequency range, and scanned with a focused laser beam
perturbation. At low temperatures, direction-dependent
nonlinearities in the reactive and resistive properties of the
resonator create photoresponse that maps out the directions of
nodes, or of bound states associated with these nodes, on the Fermi
surface of the superconductor. The method is demonstrated on the
nodal superconductor YBa$_{2}$Cu$_{3}$O$_{7-\delta}$ and the results are consistent with theoretical predictions for the bulk and surface contributions.
\end{abstract}

\maketitle

\textit{Introduction} - The Meissner effect is the spontaneous
exclusion of magnetic flux from the bulk of a superconductor and is one of the hallmarks of superconductivity. In the
presence of a magnetic field, a superconductor must invest kinetic
energy in a supercurrent flow to screen out the applied field. This
reduces the free energy difference between the superconducting and
normal states, resulting in a reduction in magnitude of the
superconducting order parameter. This in turn leads to a field- and current-dependent magnetic penetration depth, diamagnetic moment, etc. and is referred to as the nonlinear Meissner effect
(NLME). Microscopically the NLME arises when Cooper pairs at the leading edge of the current-carrying Fermi surface can de-pair into available quasi-particle states at the back-end and create a quasi-particle backflow current \cite{sem58}.  Conventional (fully gapped) superconductors show the
strongest nonlinearities near $T_{c}$, and have exponentially
suppressed nonlinear response at low temperatures, $T \ll T_{c}$.
Unconventional superconductors with nodes in the superconducting
energy gap are expected to have a strong nonlinear Meissner effect
at low temperatures, due to the nodal excitations out of the
superconducting ground state \cite{Yip1992}. In addition this
nonlinear response should be anisotropic, reflecting the locations
of nodes of the gap on the Fermi surface.

 For a $d_{x2-y2}$ gap
on a circular cut of the cylindrical Fermi surface, typical of
cuprate superconductors, theoretical treatments of the anisotropic
NLME (aNLME) predicted a $1/\sqrt{2}$ anisotropy at zero temperature
\cite{Yip1992, Xu1995}. Later, the theory was re-formulated in terms
of nonlinear microwave intermodulation response of a nodal
superconductor \cite{Dahm1997, Dahm1996, Dahm1999}, and the
temperature dependence of the NLME and its anisotropy was worked out. The dependence of the superfluid density $n_{s}$ on supercurrent density
$\vec{J_{s}}$, for sufficiently small currents, is given by
$n_{s}(T, \vec{J_{s}})=n_{s}(T)[1-b_{\Theta}(T)(J_{s}/J_{c})^{2}]$,
where $J_{c}$ is the critical current density, and $b_{\Theta}(T)$
is the nonlinearity coefficient dependent on the direction
(angle $\Theta$) of $\vec{J_{s}}$ \cite{Dahm1997}. It was found that the anisotropy in the NLME of cuprates
is weak at high temperatures, and only becomes significant for
$T/T_{c} < 0.6$ \cite{Dahm1996}. In addition, it was found that
$b_{\Theta}(T)$ is expected to grow as $1/T$ for $T/T_{c} < 0.2$
\cite{Dahm1997}, before crossing over to another temperature
dependence, depending on the purity of the material \cite{Xu1995,
Dahm1999, Li1998}.

 Early experiments to detect the aNLME of cuprates through
transverse magnetization \cite{Buan1994, Bhattacharya1999}, magnetic
penetration depth \cite{Carrington1999, Bidinosti1999,
Halterman2001}, and surface impedance \cite{Maeda1995} measurements
did not establish conclusive evidence of the effect \cite{Li1998}.
Later, sensitive nonlinear microwave measurement techniques established the existence of the NLME in cuprates from the
temperature dependence of the intermodulation power at low
temperatures \cite{Benz2001, Oates2004, Leong2005}, although the
anisotropy of the effect was not measured. Although other
thermodynamic measurements have shown evidence of an anisotropic
energy gap in various superconductors \cite{Vekhter1999, Aubin1997,
Park2003}, there are no unambiguous measurements of the aNLME in
cuprates, to our knowledge \cite{Groll2010}.  In contrast to quasiparticle transport
methods \cite{Vekhter1999, Aubin1997, Park2003}, the present method
utilizes the bulk superfluid response to identify the nodal
directions \cite{Zutic1997}.

 An additional contribution to the NLME in unconventional
superconductors arises from Andreev bound states (ABS)
\cite{Hu1994}, created, for example, on the \{110\} surfaces of a
$d_{x2-y2}$ superconductor. These states give rise to a
\textit{para}magnetic contribution to the screening
\cite{Fogelstrom1997}.  For cuprates, \{110\} interfaces occur at twin boundaries, which are formed spontaneously during epitaxial film growth.  The NLME associated with ABS has been
established by tunneling \cite{Aprili1999}, and penetration depth
measurements \cite{Walter1998, Carrington2001}, for example. Theory
by Barash, Kalenkov and Kurkij\"{a}rvi (BKK) \cite{BKK} and Zare, Dahm and Schopohl (ZDS) \cite{Zare2010} predicts an aNLME associated with
ABS having a strong temperature dependence at low temperatures,
eventually dominating that due to nodal excitations.

 Here we establish a new method to both quantitatively measure
and image the aNLME from nodes in the superconducting gap using a
novel RF resonant technique combined with laser scanning microscopy.

\textit{Experiment} - We employ a self-resonant superconducting
structure based on a thin film Archimedean spiral geometry. The
spiral has an inner diameter of 4.4 mm, an outer diameter of 6 mm,
and consists of 40 turns of nominally 10 $\mu$m width thin film
stripe with 10 $\mu$m spacing, winding continuously from the inner
to outer radii with Archimedean form (see the schematic in Fig.
1(a)). The structure has considerable inductance and capacitance,
making it a compact self-resonant meta-atom for use in
superconducting metamaterials \cite{Kurter2010, Anlage2011}. When
made from superconducting materials such as Nb and
YBa$_{2}$Cu$_{3}$O$_{7-\delta}$ (YBCO), the spiral has a fundamental
resonance in the vicinity of 75 MHz, and many overtones extending
above 1 GHz. The distribution of standing wave currents on the
spiral in the first few modes are well approximated as those of a
resonant vibrating string held fixed at both ends, and then wrapped
into a spiral, as verified by detailed laser scanning microscope
(LSM) imaging \cite{Kurter2011a, Kurter2011b, Zhuravel2012, SM}. A
unique property of the resonant spiral is the fact that the currents in the low-order modes circle the spiral many times, repeatedly sampling all parts of the in-plane Fermi surface \cite{Zhuravel2012}.

 The epitaxial and in-plane oriented YBCO films on LaAlO$_{3}$ (LAO)
were deposited by \textit{in-situ} off-axis magnetron sputtering to
a thickness of 300 nm \cite{Face1997}. Similar YBCO films on
CeO$_{2}$ buffered sapphire and on
MgO were deposited by thermal coevaporation to a thickness of 300 nm and 700 nm, respectively \cite{Theva}. The 200 nm thick Nb films were deposited on quartz
substrates at room temperature by RF sputtering \cite{Kurter2010,
Kurter2011a}. All films were patterned into spiral resonators by contact photolithography
and either wet or dry chemical etching.

 The LSM operates by scanning a focused laser spot (wavelength
640 nm) over the surface of the spiral while it is excited near its
RF resonance with excitation and pickup loops above and below the plane of the film \cite{SM}.
Details of the RF and laser excitations of the spirals
have been previously published \cite{Kurter2010, Kurter2011a,
Kurter2011b, Zhuravel2012}. The spiral develops a photoresponse due
to pair-breaking and localized heating, producing a change in its
resonant frequency and quality factor \cite{Zhuravel2006a,
Zhuravel2010}. The laser intensity is modulated at 100 kHz, and the
changes in RF transmission at a fixed frequency are phase
sensitively detected, resulting in a photoresponse (PR) signal
consisting of both a magnitude and phase. The peak laser intensity
can be varied between 150 $\mu$W and 1.6 mW. The laser spot (about
20 $\mu$m diameter) is scanned over the sample to create a
two-dimensional PR image.  To first approximation, the PR magnitude can be interpreted as the distribution of RF currents $J_{RF}^2(\rho,\Theta)$ \cite{Kurter2011b, Zhuravel2012, SM,  Zhuravel2006a, Zhuravel2010, Newman1993, Zhuravel2002}.

The aNLME is expected to create the following PR contrast
\cite{Dahm1997, Scala2011}. For a $d_{x2-y2}$ gap on a circular
Fermi surface one expects PR $\sim 1+\sin^{2}(2\Theta)$ with $\Theta
= 0$ along the Cu-O-Cu bond direction of YBCO (see Fig. 1(a)).  In addition, the PR
amplitude should vary as $1/T^{2}$ for $T/T_{c} < 0.2$ and
$J_{RF}/J_{c} < T/T_{c}$, where $J_{RF}$ is the RF current density
induced in the sample. It is also expected that the LSM PR is
proportional to RF power and proportional to the intensity of the
laser light. As deduced from BKK and ZDS, the contribution to LSM PR from
ABS should be centered on \{110\} surfaces of the material, have a
stronger temperature dependent PR $\sim 1/T^{4}$, have the opposite
sign of PR from the nodal aNLME (due to the paramagnetic nature of the ABS), and be a linear function of RF and
laser power up to RF fields on the order of a few mT
\cite{BKK, Zare2010}.

\begin{figure}\center
\includegraphics[width=3in]{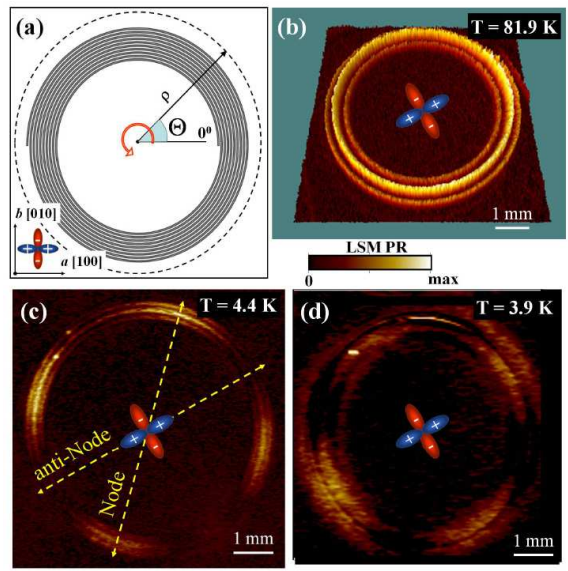}
\caption{a) Schematic diagram of spiral geometry, definition of
radial ($\rho$) and angular ($\Theta$) coordinates, and directions
of the crystallographic a- and b-directions, along with the
orientation of the d-wave gap in YBCO.  b)-d) LSM photoresponse
images of a YBCO/LAO spiral resonator in the third harmonic mode at
temperatures of b) 81.9 K, c) 4.40 K, d) 3.90 K. Also shown is the
d-wave gap orientation as determined from twin domain boundary
directions in the substrate.}
\end{figure}

\textit{Results} - Figure 1 shows LSM PR images of a single YBCO/LAO
spiral in a resonant mode at three different temperatures, 81.9 K,
4.4 K and 3.9 K. The image is of the third harmonic mode and shows
three nearly circular distinct bands of enhanced PR, corresponding
to the three half-wavelengths of current distributed over the 40
turns of superconducting wire \cite{Kurter2011a,
Kurter2011b, Zhuravel2012, SM}. One notes that the PR is isotropic in
its angular distribution around the spiral at 81.9 K (Fig. 1(b)).
However, the LSM PR image of the same mode at 4.40 K (Fig. 1(c))
shows four distinct enhancements of PR at regular angular intervals
around the spiral. As the sample is cooled further (below about 4.30
K for this particular sample), a new set of enhanced-PR features are
observed in Fig. 1(d), but now rotated $45^{o}$ relative to those at
4.40 K. Images of Nb spirals of nominally identical geometry in the
same mode do not show this anisotropic PR at any temperature below the
transition temperature. The above anisotropy and rotation behavior
is observed to hold for the YBCO/LAO spiral excited into all modes,
from the fundamental through the $8^{th}$ mode (the highest
examined).

 Figure 2 shows the radially($\rho$)-integrated PR as a function of
angle for a YBCO/LAO thin film spiral resonator taken at a
temperature of 4.66 K in the $7^{th}$ mode at 450.22 MHz, shown in
the inset. The zero-angle (Cu-O-Cu bond) direction was determined
from the visible twinning structure of the LAO substrate. Utilizing
the fact that the YBCO film was grown \textit{epitaxially} and
in-plane oriented on LAO, the observation that the PR maxima are
aligned along the linear twin domain structures in the substrate,
and the fact that the twin boundaries are aligned at $45^{o}$ to the
Cu-O-Cu bond direction \cite{Wang2008}, we deduced that the $\Theta
= 0$ direction is at $45^{o}$ to the twin domain structure of the
substrate. Figure 2 shows a fit of the data to PR$(\Theta) = 1 + A
\sin^{2}(2\Theta)$, resulting in a fit value $A = 1.05 \pm 0.02$,
close to the value of 1 expected for a simple $d_{x2-y2}$ gap on a
circular Fermi surface.

\begin{figure}\center
\includegraphics[width=3in]{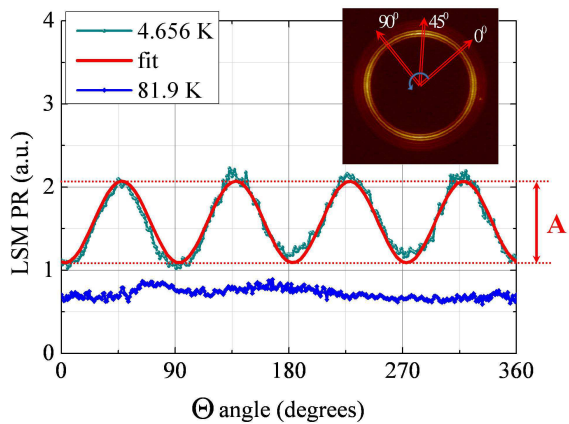}
\caption{Plot of radially-integrated and unwrapped photoresponse
(PR) vs. angle on the YBCO/LAO spiral at 81.9 K and 4.66 K, along
with a fit of the low-temperature integrated PR$(\Theta)$ to the
simple d-wave angular dependence. Inset shows PR image of
YBCO/LaAlO$_{3}$ thin film spiral resonator taken at a temperature
of 4.656 K in the 7$^{th}$ mode at 450.22 MHz.}
\end{figure}

 The PR images in the `rotated' state at lower temperatures
(Fig. 1(d)) also show 4-fold character, but are relatively diffuse
in appearance, and display dramatically stronger PR as the
temperature is lowered. To understand these properties, we examine
the temperature dependence and phase of the photoresponse along the
nodal ($\Theta = \pi/4$) and anti-nodal ($\Theta = 0$) directions of
the superconducting gap. Figure 3 shows a plot of PR magnitude vs.
temperature for nodal and anti-nodal regions of a YBCO/sapphire
spiral resonator. In the nodal direction the PR is of significant
magnitude at a temperature of 6.6 K, but decreases in magnitude as
temperature is lowered. The PR goes to near-zero levels and changes
phase by $\pi$ radians at a point that we call the `crossover
temperature.'\cite{SM}  Below this temperature the nodal PR magnitude
increases dramatically. In the anti-nodal directions the PR is found
to be smaller than the nodal PR at 6.6 K, but increases
monotonically in magnitude with decreasing temperature, and has the
same phase as the nodal PR below the crossover temperature.

\begin{figure}\center
\includegraphics[width=3in]{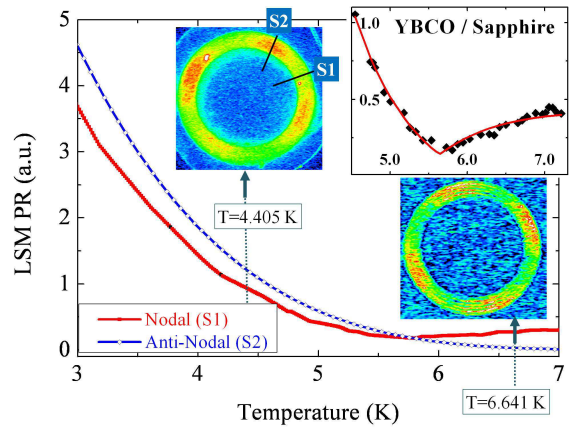}
\caption{Plot of temperature dependence of photoresponse (PR)
magnitude taken at nodal and anti-nodal locations in a YBCO/sapphire
spiral resonator. The anti-nodal PR has opposite sign to the nodal
PR above the crossover temperature (5.63 K, as deduced from the fit). Graphical inset shows
temperature dependence of nodal PR magnitude near its crossover
temperature, along with a fit to the form predicted by ZDS. Other
insets show LSM PR magnitude images of YBCO/sapphire below (4.405 K)
and above (6.641 K) its crossover temperature at 546.555 MHz. The
phase of the PR in these images differ by $\pi$, corresponding to a
sign reversal of the response.}
\end{figure}

A fit of the PR to the expected
temperature dependence of nodal PR (temperature derivative of Eq.
(16) of Ref. 25), namely PR$ = A/T^{2} - B/T^{4} + const.$ (first term due
to nodal NLME and second due to ABS NLME) \cite{Zare2010}, is shown
in the graphical inset of Fig. 3 for a YBCO/sapphire sample. The data is
consistent with the predicted temperature dependence around the
crossover temperature.

In all of these experiments we have verified that the LSM PR in the
nodal direction is proportional to both RF power and to laser
intensity. The LSM PR in the anti-nodal direction is linear at low
laser powers, but becomes strongly nonlinear at higher laser powers.
YBCO films grown on other substrates (MgO and sapphire) show the
same general behavior described above as the samples on LAO. The
spiral PR image insets of Fig. 3 show the YBCO/sapphire sample below
(4.405 K) and above (6.641 K) the crossover temperature for this
sample. We find that the PR maxima are aligned along the nodal
directions of YBCO at higher temperatures, and then show a
`rotation' by $45^{o}$ at lower temperatures,\cite{rot} although the
crossover temperature is sample dependent (approximately 4 - 5 K for
YBCO/LAO and 5 - 7 K for YBCO/MgO and YBCO/sapphire). This sample
dependence may be associated with different abundances of exposed \{110\} facets, and twinning in each sample.  YBCO films typically show twin separations less than 100 nm \cite{Streiffer} and have complex bi-directional twinning, \cite{Kastner} meaning that our focused laser spot averages over a great many twin domains and any anisotropic effects of twinning will not be manifested on the scale of the spiral images.  Note that the global nodal and anti-nodal directions are preserved in twinned structures.

\textit{Discussion} - A reorientation of specific heat oscillations with magnetic field direction (qualitatively similar to the crossover observed here) has been reported in superconductors with anisotropic order parameters \cite{An2010, Zeng2010}.  However, specific heat measures the low-energy quasiparticle density of states \cite{Vorontsov2007} while our measurements probe the superfluid response and involve no DC magnetic field or vortices. Therefore we do not believe that quasiparticle transport phenomena are the origin of the crossover observed in this paper.  On the other hand, the observed sign difference of LSM PR between the nodal
and anti-nodal directions above the crossover temperature is
consistent with the ZDS prediction for the NLME due to ABS
\cite{Zare2010}. The strong and monotonic temperature dependence of
PR observed along the \{110\} directions of the film are also
consistent with the ZDS prediction. The observed crossover
temperature (4 K to 7 K) is on the order of that predicted by ZDS,
namely $T_{c}/\kappa^{1/2}\sim$ 10 K, where $\kappa$ is the Ginzburg-Landau parameter.  Together, these results
confirm the basic predictions of BKK and ZDS for the relative contributions
of the nodal and ABS contributions to the NLME.  Note that the use
of thin film superconducting spiral resonators that expose many
different crystallographic directions, and which have a large
surface-to-volume ratio and high twin density, accentuates the contribution of ABS to the photoresponse.

 In order to verify the above interpretation of the data, a number of potential artifacts and other
interpretations of the PR images have been systematically explored.
First is the possibility of anisotropic PR being created by the
irregularly-shaped excitation and RF pickup loops. The anisotropic
PR did not change when the excitation loop was rotated relative to
the sample. In addition, the anisotropic PR pattern rotated when the
sample was rotated relative to the loops. We do not observe a
significant degree of PR anisotropy in Nb spirals measured under
similar conditions. 

 A second concern is that the PR temperature dependence arises
from a growing thermal boundary resistance, $R_{bdy} \sim 1/T^{3}$,
between the thin film superconductor and the dielectric substrate
\cite{Swartz1989}. However the strong temperature dependence is not
observed in Nb/quartz samples, eliminating most of the cryostat
components from generating this response. All YBCO samples (on LAO,
MgO and sapphire) show the strong PR temperature dependence at
low temperatures. 
Nevertheless, we do believe that a thermal blockade may set in for the YBCO/LAO samples at low temperatures ($<$ 3 K), resulting in diffuse PR
images. This interpretation was independently verified through measurements of the PR thermal relaxation time as a function of temperature.

A third concern is that the twinned structure of the LAO substrate
traps heat and channels it in specific directions, giving rise to
the observed anisotropy. However, we observe the same anisotropy on
YBCO/MgO and YBCO/sapphire films that do not have twinned
substrates. The apparent rotation of the PR by $45^{o}$ in a narrow
temperature range in all of the YBCO samples also argues against a
substrate-induced anisotropy. Next is the concern that the shape of
the substrate dictates the anisotropy of the PR in the spiral.
Numerical simulations of localized heating of a spiral on a square
substrate under conditions similar to those of our YBCO samples
showed a temperature anisotropy of only 1 part in 1000 imposed by
the shape of the substrate, whereas the observed anisotropy is on
the order of 50\%. Finally is the concern that defects in the
lithographic pattern of the spiral (namely shorts between wires or
opens) will create the anisotropic PR. We find that all YBCO
spirals, from those with near-perfect lithography to those with
defective patterns, all show the anisotropy and temperature
dependent properties described above.

\textit{Conclusions} - We have measured and imaged the anisotropic nonlinear Meissner effect in an unconventional superconductor arising from both bulk and surface mechanisms.  Our results are in detailed agreement with aNLME theoretical predictions based on both nodal and ABS contributions to the photoresponse.  This establishes a new gap spectroscopy tool that is sensitive to nodes in the superconducting gap accessed by currents flowing in the plane of the material, and can detect gap nodes directly, or indirectly through Andreev bound states, through their influence on the in-plane electrodynamics in the superconducting state.

\textit{Acknowledgements} We thank D. J. Scalapino, I. Mazin, and V. Yakovenko
for helpful discussions. The work at Maryland was supported by ONR
Grants No. N000140811058 and No. 20101144225000, the US DOE DESC
0004950, the ONR AppEl Center, Task D10 (N000140911190), and CNAM.
The work in Karlsruhe is supported the Deutsche
Forschungsgemeinschaft (DFG) and the State of Baden-W\"{u}rttemberg
through the DFG-CFN, and a NASU program on Nanostructures,
Materials and Technologies. S.M.A. acknowledges sabbatical support
from the CFN at KIT.

%




\begin{references}
\bibitem{rot} This apparent rotation occurs over a narrow temperature range because of the two competing PR contributions with opposite signs and very different temperature dependencies.
\end{references}
\end{document}